\documentstyle[preprint,aps]{revtex}

\begin{document}
\draft
\preprint{RU9687}
\title{Studies in the statistical and thermal properties of hadronic
matter under some extreme conditions}
\author{K. C. Chase, A. Z. Mekjian and P. Meenakshisundaram}
\address{Department of Physics, Rutgers University \\
Piscataway, New Jersey 08854}
\date{\today}

\maketitle

\begin{abstract}
The thermal and statistical properties of hadronic matter under some
extreme conditions are investigated using an exactly solvable
canonical ensemble model.  A unified model describing both the
fragmentation of nuclei and the thermal properties of hadronic matter
is developed.  Simple expressions are obtained for quantities such as
the hadronic equation of state, specific heat, compressibility,
entropy, and excitation energy as a function of temperature and
density.  These 
expressions encompass the fermionic aspect of nucleons, such as
degeneracy pressure and Fermi energy at low temperatures and the ideal
gas laws at high temperatures and low density.  Expressions are
developed which connect these two extremes with behavior that
resembles an ideal Bose gas with its associated Bose condensation.  In
the thermodynamic limit, an infinite cluster exists below a certain
critical condition in a manner similar to the sudden appearance of the
infinite cluster in percolation theory.
The importance of multiplicity fluctuations is discussed and some
recent data from the EOS collaboration on critical point behavior of
nuclei can be accounted for using simple expressions obtained from the 
model. 
\end{abstract}

\narrowtext


\section{Introduction}
\label{sec:Introduction}

One of the goals of nuclear physics is to characterize the thermal
properties of nuclear matter, described by its
equation of state.  Investigating heavy ion collisions offers one
experimental method of probing nuclei away from the typical low
temperature and density, and a number of experiments are underway studying
the breakup of nuclei, contributing greatly to our
understanding~\cite{Pochodzalla-J:1995prl,Kreutz-P:1993npa,Wang-S:1995prl,Kwiatkowski-K:1995prl,Handzy-DO:1995prl,Gilkes-ML:1994prl}.
Several theoretical investigations of these properties have been
instigated, 
thermal~\cite{Csernai-LP:1986prep,Goodman-AL:1984prc,Curtin-MW:1983plb,Bertsch-GF:1983plb},
statistical~\cite{Bertsch-GF:1988prep,Moretto-LG:1988ppnp,Bondorf-JP:1995prep,Gross-DHE:1990rpp,Pan-JC:1995prc}
transport~\cite{Bertsch-GF:1984prc,Gan-HH:1987prc,Aichelin-J:1991prep,Suraud-E:1992npa,Bonasera-A:1992plb} and
percolative~\cite{Campi-X:1992npa,Desbois-J:1987npa,Bauer-W:1986npa,Biro-TS:1986npa,DasGupta-S:1993plb,Elliott-JB:1994prc}
models all offering
insights into the properties of fragmenting nuclei.

In a previous set of
papers~\cite{Mekjian-AZ:1990prl,Mekjian-AZ:1990prc,Lee-SJ:1990pla,Chase-KC:1994prc1}
an exactly solvable canonical ensemble model was developed for
studying the multifragmentation of nuclei induced by high energy
collisions.  
Detailed properties of the cluster distributions such as inclusive
yields and correlations were explored.
In this paper, we study the statistical and thermal
properties of nuclei using the same model, extending results presented
in~\cite{Chase-KC:1995prl}.
This model gives a unified description of multifragmentation phenomena
and the thermodynamic properties of hadronic matter, such as its equation
of state and the nuclear compressibility.  Simple expressions for
these quantities are obtained which extrapolate or connect between
a low temperature nearly degenerate Fermi gas to a high temperature
ideal Maxwell-Boltzmann gas of nucleons.  
Importantly, the effects of the fragmentation degrees of freedom on
the thermodynamics is manifestly included.
Questions related to very high temperatures and particle and
anti-particle production will be developed in a subsequent paper.  

\section{Model studies}
\label{sec:model-studies}

Given the partition function $Z_{A}(V, T)$ for a system of $A$
nucleons at a given volume $V$ and temperature $T$ we can determine
the thermodynamic functions by taking various partial derivatives,
i.e. the internal energy and pressure are given by
$U = k_{B} T^{2} \left( {\partial \over \partial T} \ln Z_{A} \right)_{V}$,
$P = k_{B} T \left( {\partial \over \partial V} \ln Z_{A} \right)_{T}$.
Additional partial derivatives lead to such important quantities as the
specific heat and isothermal compressibility,
$C_{V} = \left( {\partial U \over \partial T} \right)_{V}$,
${1 \over \kappa_{T}} = -\left( V {\partial P \over \partial V} \right)_{T}$,
and thermodynamic potentials such as the entropy and Gibbs free energy
can be related to these quantities, 
$S = U/T + k_{B} \ln Z_{A}$,
$G = -k_{B} T \ln Z_{A} + P V$.
Our interest of course is not in the partition function itself, but
rather in these derived functions, and as such the partition function is
merely a means to that end.

This section constructs a partition function for fragmenting nuclear
matter.  It does so by investigating the thermodynamic functions at
both low and high temperatures and densities.  At low temperature/high
density, the nucleons
coalesce into a single nucleus which can be modeled
as a degenerate Fermi gas with Skyrme interactions.  At high
temperature/low density,
the nucleons evaporate into a Maxwell-Boltzmann gas of individual
nucleons.  With these results firmly in mind, a
statistical fragmentation model is introduced whose parameters depend
on the thermodynamics in a way consistent with the high and low temperature
limits, yet allowing more complicated fragmentation patterns at
intermediate temperatures.
RHS always diverges in the infinite limit
The remaining freedom in the model is eliminated by assuming
that the cluster yields follow a power law.  Interestingly these
minimal assumptions lead to a model surprisingly rich in physics as we
shall see in the next section.

\subsection{Low temperature behavior and ideal Fermi gas laws}
\label{sec:low-temp}

At zero temperature, the nucleus can be treated as a
degenerate Fermi gas, with Fermi momentum $p_{F}$ given by
\begin{equation}
g_{S,I} {4 \over 3} \pi p_{F}^{3} = h^{3} \rho \;,
\end{equation}
where $g_{S,I} = 4$ is the spin-isospin degeneracy factor and
$\rho = A/V$ is the density.  At normal
density $\rho_{0} = A/V_{0} = 0.17$ nucleons fm$^{-3}$, nucleons are
non-relativistic and the Fermi energy and momentum are related by
$\varepsilon_{F} = p_{F}^{2}/2m$.  From the above expression for the
Fermi momentum, we see that $\varepsilon_{F} \propto \rho^{2/3}$.
At low temperatures, the total kinetic energy for such a Fermi gas
is given by~\cite{Huang-K:1987-stat-mech}
\begin{equation}
E_{K} = {3 \over 5} A \varepsilon_{F} 
  \left( 1 + {5 \pi^{2} \over 12} \left(
    {k_{B} T \over \varepsilon_{F}} \right)^{2} \right) \;.
\label{eq:low-kinetic}
\end{equation}
However, the nucleons do interact.
Interactions between the nucleons can be taken into account by using a 
density dependent Skyrme
interaction~\cite{Vautherin-D:1972prc,Skyrme-THR:1959nucp}.  
In this approach, the internal energy of a nucleus is given by
\begin{equation}
U_{{\rm LT}} = E_{K} - A \left( a_{0} {\rho \over \rho_{0}} -
   a_{3} \left( {\rho \over \rho_{0}} \right)^{1+\sigma} \right) \;,
\end{equation}
where $a_{0}$ and $a_{3}$ are Skyrme parameters (for simplicity,
temperature independent) determined by fixing
$a_{V}(\rho) = -U_{{\rm LT}}(T=0)/A$ the binding energy per particle at zero
temperature
\begin{equation}
a_{V}(\rho) 
  = -{3 \over 5} \epsilon_{F}(\rho) +
         a_{0} \left( {\rho \over \rho_{0}} \right) -
	a_{3} \left( {\rho \over \rho_{0}} \right)^{1+\sigma}
\end{equation}
to the empirical value for $a_{V}$
at $\rho = \rho_{0}$ (about 8.0 MeV/nucleon at zero temperature)
and having this as the maximum of $a_{V}$.  The
density dependent repulsive term should appear with a higher power of
$\rho$ than the attractive part, so that the nucleus does not
collapse, and thus requires $\sigma > 0$.
Commonly used values for $\sigma$ are $\sigma = 1$
(three body interaction) and $\sigma = 2/3$ (finite range
term)~\cite{Jaqaman-HR:1983prc}.

Having specified the internal energy, we can now determine any other
thermodynamic function.  The specific heat
$C_{V} = \left({\partial U \over \partial T} \right)_{V}$, and
entropy $S = \int^{T} C_{V}/T'\ dT'$
are therefore
\begin{equation}
\left( {C_{V} \over k_{B}} \right)_{{\rm LT}} = 
\left( {S \over k_{B}} \right)_{{\rm LT}} =
  A {\pi^{2} \over 2} 
  \left( {k_{B} T \over \varepsilon_{F}} \right) \;.
\label{eq:entropy-lowt}
\end{equation}
The partition function can now be determined from
$Z_{A} = \exp \{ (ST-U)/k_{B}T \}$, which gives
\begin{equation}
Z_{A} = y^{A} \;,
\end{equation}
where
\begin{equation}
y = \exp \left\{
    {a_{V} \over k_{B}T } 
  + {k_{B} T \over \varepsilon_{0}} \right\} \;,
  \label{eq:y-nocutoff}
\end{equation}
and $\varepsilon_{0} = {4 \over \pi^{2}} \varepsilon_{F}$.
>From this partition function it is easy to determine the
pressure and compressibility.
\begin{eqnarray}
P_{{\rm LT}} 
  & = & {2 \over 5} \rho \varepsilon_{F}(\rho) 
        \left( 1 + {5 \pi^{2} \over 12} 
        \left( {k_{B} T \over \varepsilon_{F}} \right)^{2} \right) 
        \nonumber \\
    &&- a_{0} \rho {\rho \over \rho_{0}} +
            (1+\sigma) a_{3} \rho 
            \left( {\rho \over \rho_{0}} \right)^{1+\sigma}
        \;, \\ 
\left( {1 \over \kappa_{T}} \right)_{{\rm LT}}
  & = & {2 \over 3} \rho \varepsilon_{F}
         \left( 1 + {\pi^{2} \over 12} 
         \left( {k_{B} T \over \varepsilon_{F}} \right)^{2} \right) 
         \nonumber \\
    && - 2a_{0} {\rho^{2} \over \rho_{0}} +
          (1+\sigma)(2+\sigma) a_{3} \rho
           \left( {\rho \over \rho_{0}} \right)^{1+\sigma} .
\end{eqnarray}

\subsection{High temperature, low density ideal Maxwell-Boltzmann gas laws}
\label{sec:high-temp}

At high temperature and/or low density, but below meson and
particle-antiparticle production thresholds, the hadronic properties
are that of an ideal Maxwell-Boltzmann gas, with
\begin{eqnarray}
U_{{\rm HT}}  
  & = & {3 \over 2} A k_{B} T \;, \\
\left( {C_{V} \over k_{B}} \right)_{{\rm HT}} 
  & = & {3 \over 2} A \;, \\
P_{{\rm HT}} 
  & = & \rho k_{B} T \;, \\
\left( {1 \over \kappa_{T}} \right)_{{\rm HT}}
  & = & \rho k_{B} T \;, \\
\left( {S \over k_{B}} \right)_{{\rm HT}}
  & = & A \ln \left\{ e^{5/2} 
        {V \over A \lambda_{T}^{3}} g_{S,I} \right\} \;,
\label{eq:Sackur-Tetrode}
\end{eqnarray}
where the entropy $S$ is given by the Sackur-Tetrode law and
$\lambda_{T} = h/\sqrt{2 \pi m k_{B} T}$.
We expect that nuclei vaporize into individual nucleons at 
$k_{B} T \gg a_{V}$, the binding energy per particle.
The partition function consistent with these thermodynamic functions
is $Z_{A} = x^{A}/A!$, where $x = V/\lambda_{T}^{3}$.

\subsection{Hadronic matter at moderate temperatures and densities and
ideal Bose gas-like structure}
\label{sec:inter-temp}

In the region between the ideal Fermi gas limit and the ideal
Maxwell-Boltzmann gas limit, the effects of the fragmentation of the
initial nucleus must be taken into account when considering
thermodynamic issues.  To account for these effects,
we employ a model developed initially to study multifragmentation
phenomena~\cite{Lee-SJ:1992prc1}.   This model has the correct
high and low temperature 
limits, as determined in the previous sections, but has features
similar to that of a Bose gas in the intermediate range.

To describe the fragmentation of a nucleus into all possible modes of
breakup, a weight is given to each possibility.  The weight chosen is
\begin{equation}
W({\bf n}) 
   = \prod_{k \ge 1} {x_{k}^{n_{k}} \over n_{k}!}
   = \prod_{k \ge 1} {1 \over n_{k}!} 
     \left( {x y^{k-1} \over \beta_{k}} \right)^{n_{k}} \;,
     \label{eq:weight}
\end{equation}
where $x$ and $y$ are functions of the thermodynamic variables 
$(V, T)$, $\beta_{k}$ is the cluster size dependence
of the weight and ${\bf n} = (n_{1}, n_{2}, \ldots)$ is the
fragmentation vector, with $n_{k}$ the number of fragments with $k$
nucleons such that $\sum_{k} k n_{k} = A$.  With this choice of weight
the canonical partition function $Z_{A} = \sum_{{\bf n}} W({\bf n})$
is a polynomial in $x, y$, given by
\begin{equation}
Z_{A}(x, y) 
  = \sum_{m=1}^{A} Z_{A}^{(m)}(\vec{\beta}) x^{m} y^{A-m} \;,
    \label{eq:pf-decomposed}
\end{equation}
with $m = \sum_{j} n_{j}$ the multiplicity.
Given the partition function, all thermodynamic properties of the
model can be obtained, as well as ensemble averages.
For example, the mean number of clusters of size $k$ is
\begin{equation}
\langle n_{k} \rangle = {x y^{k-1} \over \beta_{k}} 
  {Z_{A-k}(x, y) \over Z_{A}(x, y)} \;.
\end{equation}
The partition functions themselves can be obtained using a recursive
procedure defined by the constraint 
$\sum_{k} k \langle n_{k} \rangle = A$, which gives
\begin{equation}
Z_{A}(x, y) = {1 \over A} \sum_{k=1}^{A} k 
  {x y^{k-1} \over \beta_{k}} Z_{A-k}(x, y) \;,
\end{equation}
where $Z_{0}(x, y) = 1$.  The whole procedure is easily implemented
by computer.

The parameters $x, y$ determine the thermodynamic aspects of the
models and as such need to be determined.  They are most easily
determined by considering the partition function at high and low
multiplicity.  When the 
multiplicity is large $\langle m \rangle \approx A$, and 
$Z_{A} \approx Z_{A}^{(A)} x^{A}$. 
Clearly this is the high temperature limit, and $x$ is simply 
the ideal gas $x$ introduced in section~\ref{sec:high-temp},
given by
\begin{equation}
x = {V \over \lambda_{T}^{d}}
\end{equation}
and involves the volume of the system $V$, the thermal
wavelength $\lambda_{T}$, and the dimensionality of the system $d$.
The term $x^{m}$ in the weight arises from the thermal motion of each
fragment.  Since the overall motion is zero, this should be replaced
by $x^{m-1}$ to reflect the center of momentum constraint.
When the multiplicity is small $\langle m \rangle \approx 1$, and 
$Z_{A} \approx Z_{A}^{(1)} x y^{A-1}$.  Assuming that the $x$ parameter is
removed by the conservation of momentum constraint describe above,
this shows $y$ to be given by Eq.~(\ref{eq:y-nocutoff}).  A correction
to the internal excitations however needs to be made at high
temperatures to reflect their finite lifetimes.  Koonin and
Randrup~\cite{Koonin-SE:1987npa} argue that a simple cutoff is
effective, which results in
\begin{equation}
y = \exp \left\{ {a_{V} \over k_{B} T} + 
          {k_{B} T \over \varepsilon_{0}} 
          {T_{0} \over T+T_{0}}
        \right\} \;.
    \label{eq:y-withcutoff}
\end{equation}
Here $a_{V}$ is the binding energy per particle, $\varepsilon_{0}$ is
the level spacing parameter (related to the Fermi energy) and $T_{0}$
is the cutoff temperature. 
So the term $y^{A-m}$ in the weight arises from the binding and internal
excitations of each fragment.

The only parameter now to determine is $\beta_{k}$.  Since the mass
yield distributions are often well represented by a power
law~\cite{Finn-JE:1982prl}, we choose $\beta_{k}$ simply to reproduce
this important experimental fact.  If $\beta_{k} = k^{\tau}$, then
in the grand canonical limit $\langle n_{k} \rangle \sim k^{-\tau}$.
Typically $\tau$ is found between two and three, and we have chosen
$\tau = 2.5$ as representative for calculations made in this paper.

\section{Thermodynamic properties of hadronic matter}
\label{sec:thermodynamics}

In this section we consider the various fundamental thermodynamic
functions of fragmenting nuclei, namely 
the pressure, compressibility and Gibbs free energy
(section~\ref{sec:pressure}),
the internal energy and specific heat (section~\ref{sec:energy}),
and the entropy (section~\ref{sec:entropy}).
Interesting features of the functions are explored, although a
detailed examination of their origin is postponed till the next section.
A brief accounting of
such properties has already been given~\cite{Chase-KC:1995prl}.

Since the partition function can be expressed succinctly as a function
of $x, y$, derivatives of $\ln Z_{A}$ with respect to these variables
will appear and reappear in computations of thermodynamic functions.
Fortunately for this model, derivatives with respect to $x, y$ yield
expectation values of the multiplicity, i.e.
$x {\partial \over \partial x} \ln Z_{A} = \langle m \rangle$,
$y {\partial \over \partial y} \ln Z_{A} = A-\langle m \rangle$,
$\left( x {\partial \over \partial x} \right)^{2} \ln Z_{A} =
 \left( y {\partial \over \partial y} \right)^{2} \ln Z_{A} =
 {\rm Var}(m) = \langle m^{2} \rangle - \langle m \rangle^{2}$.
A general derivation of these identities is included in
appendix~\ref{sec:appendix}.  

\subsection{The hadronic equation of state and a van der Waals like
structure}
\label{sec:pressure}

The hadronic equation of state is of much interest since the pressure
and incompressibility reflect the behavior of matter at fixed
temperatures and varying volumes.  A simple
expression can be obtained for this equation of state by using the
$P = k_{B} T {\partial \over \partial V} \ln Z_{A}$ 
and the partition function of Eq.~(\ref{eq:pf-decomposed}) as
determined from the weight in Eq.~(\ref{eq:weight}), giving
\begin{equation}
P = {\langle m \rangle \over A} P_{{\rm HT}} + 
  \left(1 - {\langle m \rangle \over A}\right) P'_{{\rm LT}} \;,
  \label{eq:eos}
\end{equation}
where 
$\langle m \rangle = \sum_{k} \langle n_{k} \rangle$ is the mean
multiplicity, $P_{{\rm LT}}$, $P_{{\rm HT}}$ are the low and high
temperature limits defined in
sections~\ref{sec:low-temp},~\ref{sec:high-temp} and
$P'_{{\rm LT}} = P_{{\rm LT}} - {2 \over 3} \rho
\varepsilon_{0} \left( {k_{B}T \over \varepsilon_{0}} \right)^{2} 
{T \over T+T_{0}}$ is simply the cutoff corrected low temperature
pressure.  The above equation shows the importance of the mean
multiplicity in the behavior of $P$.
For low $x/y$, $\langle m \rangle \approx 1$, and the pressure reduces
to the Fermi gas and Skyrme pressure
$P_{{\rm LT}}$.  At high
$x/y$, $\langle m \rangle \approx A$ and the ideal Maxwell-Boltzmann gas
pressure $P_{{\rm HT}}$ in $P$ dominates.  
Thus the result of Eq.~(\ref{eq:eos})
connects the two extremes in a simple analytic way.  
Figure~\ref{fig:pressure}(a) plots the pressure
for several temperatures for $\beta_{k} = k^{5/2}$.

The low temperature component of the pressure is quite interesting.
It is responsible for the pressure rising as the volume increases over
a range of volumes at low enough temperatures, a feature
characteristic of a van der Waals 
gas or a liquid-gas phase instability~\cite{Goodman-AL:1984prc}.  It
arises due to the nuclear interactions.
To see this phase instability more clearly, we can calculate the 
isothermal compressibility from the equation of state, since
${1 \over \kappa_{T}} = (\rho {\partial P \over \partial \rho})_{T}$, a little
effort gives us
\begin{eqnarray}
{1 \over \kappa_{T}} 
  & = & {\langle m \rangle \over A} P_{{\rm HT}} 
      + \left (1 - {\langle m \rangle \over A} \right) 
           \rho {\partial P'_{{\rm LT}} \over \partial \rho}
        \nonumber \\
  &&- {{\rm Var}(m) \over A} 
      {(P_{{\rm HT}} - P'_{{\rm LT}})^{2} \over P_{{\rm HT}}} \;.
\end{eqnarray}
At large values of $x/y$, $\langle m \rangle \approx A$, 
${\rm Var}(m) \approx 0$ and 
${1 \over \kappa_{T}} \rightarrow P_{{\rm HT}}$, the ideal gas limit.
At low $x/y$, $\langle m \rangle \approx 1$, ${\rm Var}(m) \approx 0$
and ${1 \over \kappa_{T}} \rightarrow P'_{{\rm LT}}$, an ideal Fermi
gas with Skyrme interaction limit.  In between, there is an effect due to the
fluctuation in the mean number of clusters.  
The behavior of the compressibility is shown in
Fig.~\ref{fig:pressure}(b).  From the figure, the phase instability
appears first at $\rho = 0.5 \rho_{0}$, $T = 24$ MeV, the point where
${1 \over \kappa_{T}}$ and its derivative are both zero.

Another method of seeing this instability is to consider the Gibbs
free energy $G = -k_{B}T \ln Z_{A} + PV$.  
Figure~\ref{fig:gibbs} shows that $G$ vs. $P$ is multivalued below a
critical temperature, a clear indication of a phase instability.

In nuclear physics, one usually refers to the
incompressibility $\kappa = {9 \over \rho \kappa_{T}}$, which if
$\varepsilon_{F} \approx 37$ MeV gives at $T=0$
\begin{equation}
\kappa = 96 + 144 \sigma {\rm MeV}
\end{equation}
For $\sigma = 1$, this model $\kappa = 240$ MeV, consistent with the
usually quoted value for the incompressibility.

\subsection{Internal energy, specific heat and the ideal Bose gas}
\label{sec:energy}

The energy and specific heat are important quantities also, especially
when the volume is fixed and the temperature is changing.
The internal energy decomposes into low and high temperature
components just as the pressure does, namely
\begin{equation}
U = {\langle m \rangle \over A} U_{{\rm HT}} + 
  \left(1 - {\langle m \rangle \over A}\right) U'_{{\rm LT}} \;,
\end{equation}
where 
$U'_{{\rm LT}} = U_{{\rm LT}} - A
\varepsilon_{0} \left( {k_{B}T \over \varepsilon_{0}} \right)^{2} 
{T(T+2T_{0}) \over (T+T_{0})^{2}}$ is simply the cutoff corrected low
temperature energy.  
Figure~\ref{fig:spec-heat}(a) plots this behavior for a number of
densities, which reveals little except for the monotonic increasing
nature of the energy.

The specific heat on the other hand is more revealing.
Using $C_{V} = \left( {\partial U \over \partial T} \right)_{V}$, 
one arrives at
\begin{eqnarray}
{C_{V} \over k_{B}} = &&
  \langle m \rangle {d \over 2}
  + (A - \langle m \rangle) {2 k_{B} T \over \varepsilon_{0}}
    \left( {T_{0} \over T+T_{0}} \right)^{3}
    \nonumber \\
  &&+ {\rm Var}(m)
    \left({d \over 2} + {a_{V} \over k_{B} T}
      - {k_{B} T \over \varepsilon_{0}}
        \left( {T_{0} \over T+T_{0}} \right)^{2}
    \right)^{2} \;,
    \nonumber
\end{eqnarray}
and Fig.~\ref{fig:spec-heat}(b) illustrates the behavior of 
$C_{V}$ for $d=3$, $\beta_{k} = k^{5/2}$.
The ideal gas limit is seen in the first term 
$\langle m \rangle {d \over 2}$.
The low temperature Fermi gas result is contained in the second term
$(A-\langle m \rangle) (2 k_{B} T/\varepsilon_{0})$.
The last term involving the multiplicity fluctuations gives rise to a
peak in $C_{V}/k_{B}$ when $d > 2$.  

The variance induced peak is quite interesting.  As we shall see in
section~\ref{sec:bose-einstein}, it is analogous to
condensation in an ideal Bose-Einstein gas, a critical transition
which has been extensively studied.

\subsection{The entropy of nuclei}
\label{sec:entropy}

The entropy of fragmenting nuclei can be obtained from the relation 
$S = {\partial \over \partial T} k_{B} T \ln Z_{A}$ which gives
\begin{eqnarray}
{S \over k_{B}} 
  & = & \ln Z_{A} + \langle m \rangle {d \over 2}
        \nonumber \\
    &&+ (A - \langle m \rangle) \left({-a_{V} \over k_{B}T} +
       {k_{B} T \over \varepsilon_{0}} 
      \left( {T_{0} \over T+T_{0}} \right)^{2} \right) \;.
\end{eqnarray}
When $\langle m \rangle \approx 1$,
the entropy is that of a nearly degenerate Fermi gas of $A$ nucleons
confined to a cluster of size $A$, plus a contribution from the motion
of this cluster which is coupled to a heat bath in the canonical
ensemble.  The nearly degenerate Fermi gas entropy is given by
Eq.~(\ref{eq:entropy-lowt}) where cutoff effects are excluded.
Including it gets
\begin{equation}
{S \over k_{B}} = A {\pi^{2} \over 2} 
  \left( {k_{B} T \over \varepsilon_{0}} \right) 
   \left( 1 - {T \over 2(T+T_{0})} \right) \;.
\end{equation}
The cutoff correction is of the same order as the next higher order 
correction.  When $\langle m \rangle \approx A$, $S$ reduces to the
Sackur-Tetrode law of Eq.~(\ref{eq:Sackur-Tetrode}).  In the
intermediate regime, their is a critical point where a latent heat
must be overcome, causing a change in volume.

Figure~\ref{fig:entropy} is a plot of $S/k_{B}A$ vs. $k_{B}T$.
In the plot, the contribution from the thermal motion of the largest
cluster has been subtracted out so that $S \rightarrow 0$ as 
$\langle m \rangle \rightarrow 1$.  This is equivalent to requiring
the total momentum of the system to be zero, a correction consistently
made throughout these calculations.

\section{Critical point behavior and parallels with other models}
\label{sec:critical-point}

The thermodynamic functions strongly suggest a critical point, which
we determine explicitly in this section.
To start, let us work out some properties of the cluster distributions
in the grand canonical ensemble.
Here the partition function is given by 
${\cal Z} = \sum_{A} z'^{A} Z_{A}(x, y) = \exp \sum_{k} (x/y) z^{k}/k^{\tau}$
where $\beta_{k} = k^{\tau}$ and $z = z' y = e^{\mu/k_{B} T}$ is the
chemical potential.  In this limit $\langle n_{k} \rangle$ is simply
\begin{equation}
\langle n_{k} \rangle 
  = {x \over y} {z^{k} \over k^{\tau}} \;.
\end{equation}
Since $A = \sum_{k} k \langle n_{k} \rangle$, we have
\begin{equation}
{A \over x'} 
  = \sum_{k=1}^{A} z^{k} k^{1-\tau} = g_{\tau-1}(z) \;,
    \label{eq:cons-of-mass}
\end{equation}
with $x' = x/y$, $x, y$ as defined earlier and 
$g_{n}(z) = \sum_{k>0} k^{-n} z^{k}$.
Notice that $g_{n-1}(z) = z {\partial g_{n}/\partial z}$ and that 
for $z \approx 1$,
\begin{equation}
g_{n}(z) = \nu^{n-1} \Gamma(1-n) + 
  \sum_{k \ge 0} \zeta(n-k) (-1)^{k} {\nu^{k} \over k!} \;,
\end{equation}
where $\nu = - \ln z$, a result due to
London~\cite{London-F:1954-superfluids2}.

The LHS of Eq.~(\ref{eq:cons-of-mass}) is always finite in the
thermodynamic limit.  If $z > 1$, the 
RHS always diverges in the infinite limit $A \rightarrow \infty$, and the
equation cannot hold.  At $z=1$, $\sum_{k} k^{1-\tau} = \zeta(\tau-1)$,
the Riemann zeta function, which is finite only if $\tau > 2$.  If
$A/x' \le \zeta(\tau-1)$ then a $z \le 1$ can be found
that satisfies the last constraint equation including all terms in the
sum.  Otherwise, the grand canonical expression breaks down, which
defines a critical point
\begin{equation}
{A \over x'_{c}} = \zeta(\tau-1) \;.
\end{equation}
The failure of the grand canonical ensemble is due to the appearance
of an infinite cluster, i.e. of size $\alpha A$ where $\alpha$ is
nonzero even as $A$ tends to infinity.  More specifically,
when $A/x'$ exceeds $\zeta(\tau-1)$, an infinite cluster exists,
absorbing enough mass so that the remaining mass is distributed 
grand canonically with $z=1$.  If
$A/x'$ is less that $\zeta(\tau-1)$, an infinite cluster does not
exist.

As we have already seen, the multiplicity and its fluctuations 
play an important role in the thermodynamic functions.
So how does the critical point affect the multiplicity and its fluctuations?
The multiplicity $\langle m \rangle = \sum_{k} \langle n_{k} \rangle$
is given simply by 
\begin{equation}
{\langle m \rangle \over A} = 
  \left\{
    \begin{array}{ll}
      {x' g_{\tau}(1) \over A} & x' < x'_{c} \\    
      {x' g_{\tau}(z) \over A} & x' > x'_{c}
    \end{array}
  \right. \;.
\label{eq:m-gc}
\end{equation}
The variance of the multiplicity 
${\rm Var}(m) = x {\partial \over \partial x} \langle m \rangle$
is given by
\begin{equation}
{{\rm Var}(m) \over A} = 
  \left\{
    \begin{array}{ll}
      {x' g_{\tau}(1) \over A} & x' < x'_{c} \\
      {x' g_{\tau}(z) \over A} - {g_{\tau-1}(z) \over g_{\tau-2}(z)} 
            & x' > x'_{c}
    \end{array}
  \right. \;.
\label{eq:varm-gc}
\end{equation}
Figure~\ref{fig:mult}(a),(b) plots 
${\rm Var}(m) = \langle m^{2} \rangle - \langle m \rangle^{2}$. 
vs. $x'$ and $\langle m \rangle$ respectively for both the canonical
and grand canonical solutions.
The discontinuity in the slope of ${\rm Var}(m)$ at the critical point
is apparent from the figure.
The behavior of ${\rm Var}(m)$ vs. $x$ shows a cusp-like behavior.
For finite systems, ${\rm Var}(m)$ has a rounded peak.  As we shall
see in the next section, this behavior is related to the ideal
Bose-Einstein gas condensation.

Is this critical behavior consistent with the experimental situation
in nuclear collisions?
At the critical point $x = x_{c}$, where ${\rm Var}(m)$ has a 
cusp, $\langle m \rangle = \langle m \rangle_{c}$ with
\begin{equation}
{\langle m \rangle_{c} \over A} = {\zeta(\tau) \over \zeta(\tau-1)}
\end{equation}
The EOS collaboration experiment determined that for gold
multifragmentation, the charge multiplicity 
$\langle m \rangle_{c} = 26 \pm 1$.  Using the above expression and   
$Z=79$ for $A$, results in a critical exponent 
$\tau = 2.262 \pm 0.013$.
Using a percolation theoretic analysis,
they arrived at a somewhat different $\tau = 2.14 \pm 0.06$.
The two results are sufficiently close to suggest that the connection
between the critical multiplicity and critical exponent
$\tau$~\cite{Chase-KC:1996plb} may not 
differ greatly from this simple model.

\subsection{Ideal Bose-Einstein gas laws}
\label{sec:bose-einstein}

Let us now compare these results with the ideal Bose gas of $A$ particles.
The results can be found in Huang~\cite{Huang-K:1987-stat-mech} for
the case $d=3$ dimensions, and generalizing to arbitrary dimensions is
fairly straightforward.  

In the grand canonical ensemble, the ideal Bose gas of $A$ particles
moving in a volume $V$ at a temperature $T$ has a fugacity 
$z = \exp \{ \mu/k_{B}T \}$ determined by
\begin{equation}
{A \over V} = {1 \over \lambda_{T}^{d}} g_{d/2}(z) + 
  {1 \over V} {z \over 1-z} \;.
\end{equation}
The critical temperature for a particular density and temperature
occurs when $z \rightarrow 1$, $V \rightarrow \infty$, which implies
\begin{equation}
\rho_{c} = {A \over V} = {1 \over \lambda_{c}^{d}} g_{d/2}(1)
\end{equation}
since above the critical point, the second term can be neglected in
the infinite volume limit.  Recalling that $x = V/\lambda_{T}^{d}$,
this condition is identical to the fragmentation case if we replace
$x$ by $x'$ and $\tau$ by $1+d/2$.

The energy, specific heat, pressure and incompressibility above and
below the critical point are given by
\begin{eqnarray}
{U \over k_{B} T} 
  & = & \left\{
  \begin{array}{ll}
    {d \over 2} {V \over \lambda^{d}} g_{1+d/2}(z) & T>T_{c} \\
    {d \over 2} {V \over \lambda^{d}} g_{1+d/2}(1) & T<T_{c}
  \end{array} \right. \;, \\
{C_{V} \over k_{B}} 
  & = & \left\{
  \begin{array}{ll}
    {d(d+2) \over 4} {V \over \lambda^{d}} g_{1+d/2}(z) -
      {d^2 \over 4} {g_{d/2}(z) \over g_{d/2-1}(z)} & T>T_{c} \\
    {d(d+2) \over 4} {V \over \lambda^{d}} g_{1+d/2}(1) & T<T_{c} 
  \end{array} \right., \\
{P} 
  & = & \left\{
  \begin{array}{ll}
    {k_{B} T \over \lambda^{d}} g_{1+d/2}(z) & T > T_{c} \\
    {k_{B} T \over \lambda^{d}} g_{1+d/2}(1) & T < T_{c}
  \end{array} \right. \;, \\
{1 \over \kappa_{T}}
  & = & \left\{
  \begin{array}{ll}
      {k_{B}T \over \lambda^{d}} {g_{d/2}^{2}(z) \over g_{d/2-1}(z)} & T > T_{c} \\
    0 & T < T_{c}
  \end{array} \right. \;.
\end{eqnarray}
Both above and below the critical point these results agree with the
fragmentation case $y=1$, $\tau = 1+d/2$ if we apply
Eqs.~(\ref{eq:m-gc})~and~(\ref{eq:varm-gc}).

This agreement can be understood by noting that the
weight given to the ideal Bose gas in $d$ dimensions in a Mayer
cluster expansion~\cite{Huang-K:1987-stat-mech} is given by
Eq.~(\ref{eq:weight}) with parameter vector
\begin{equation}
x_{k} = {x \over k^{1+d/2}} + {1 \over k}
\end{equation}
At large $x$ (i.e above the critical point) the second term is
negligible and the model reverts to the 
one considered here.  Below the critical point, the large cluster
formed contributes little to the thermodynamic functions, allowing
agreement to continue.

The nature of the transition is first order as expected for a
Bose-Einstein like condensation.  The particles which have accumulated
into the zero momentum mode contribute no pressure and have no volume or
entropy in the thermodynamic limit.  The pressure due to the balance
of particles is constant for a given temperature (i.e. independent of volume).
A latent heat per particle of
\begin{equation}
L = \left( {g_{1+d/2}(1) \over g_{d/2}(1)} \right) 
    {d+2 \over 2} k_{B} T \;,
\label{eq:latent-heat}
\end{equation}
accompanies the transition and of course is the heat $T \Delta S$
which must be spent in moving a particle from the vapor phase which has
entropy to the condensed phase which has no entropy.
The latent heat of Eq.~(\ref{eq:latent-heat}) is the result of the
change in volume along the constant pressure Maxwell line at a fixed
$T$.  The volume change is
\begin{equation}
V_{c} = {A \over \lambda_{T}^{d}} {1 \over g_{d/2}(1)} \;.
\end{equation}
The entropy change can be obtained from the Clausius-Clapeyron equation
$dP_{c}/dT = S_{c}/V_{c}$ to give
\begin{eqnarray}
{S_{c} \over k_{B}}
  & = & {d+2 \over 2} A {g_{1+d/2}(1) \over g_{d/2}(1)} \;,
\end{eqnarray}
which is consistent with $L = k_{B} T S_{c}$ as expected.
The fraction of the number of particles in the Bose condensate (zero
momentum mode) is
\begin{equation}
{N_{c} \over N} = 1 - \left( {T \over T_{C}} \right)^{d/2} \;,
\end{equation}
for $T \le T_{c}$.   

In nuclear fragmentation, this condensation is also present, but the
clusterization occurs in real space instead of momentum space.  Rather
than the number of particles in the zero momentum mode signaling a
phase transition, the number of particles in the largest cluster plays
the same role.

\subsection{Critical point behavior and percolation theory}

Percolation models~\cite{Stauffer-D:1992-Intro-perc} have been used to
describe nuclear fragmentation 
with percolation clusters corresponding to nuclear clusters arising
from the collision process.  The percolation cluster distribution is
given in terms of the percolation probability $p$ and percolation
threshold probability $p_{c}$ above which an infinite cluster exists.
Specifically,
\begin{equation}
\langle n_{k} \rangle = {1 \over k^{\tau}} f((p-p_{c})/p_{c}) \;,
\end{equation}
where $f(x)$ is a scaling function.
Percolation models have also suggested useful methods of analyzing nuclear
fragmentation data such as Campi plots~\cite{Chase-KC:1995prc}.

The model in section~\ref{sec:model-studies} has some features that
are similar with the percolation model.  First, only one parameter
$x' = x/y$ describes the distribution, the analog of $p$ in
percolation theory.
Second, below a critical point $x'_{c}$, the fraction of particles 
found in the largest cluster is finite, even if the number of
particles tends to infinity, a situation familiar from percolation
theory where for $p > p_{c}$ an infinite cluster exists, 
while for $p<p_{c}$ it can not exist.  
For the model of section~\ref{sec:model-studies} this can be seen by
analogy with the Bose condensate, where a finite fraction of
particles end up in the 
zero momentum state.  In this nuclear fragmentation model, a finite
fraction of the mass ends up in the largest cluster.
Specifically if $k_{{\rm max}}$ is
the size of the largest cluster in each event, then in 
the infinite $A$ limit
\begin{equation}
{\langle k_{{\rm max}} \rangle \over A} 
  = \left\{ \begin{array}{ll}
      1 - {x \over x_{c}} 
    = 1 - {\langle m \rangle \over \langle m \rangle_{c}} & x < x_{c} \\
                     0 & x > x_{c} 
    \end{array} \right. \;.
\end{equation}
This threshold behavior of $\langle k_{{\rm max}} \rangle/A$ as
expected exactly parallels Bose-Einstein condensation.  
Figure~\ref{fig:kmax} shows the behavior of 
$\langle k_{{\rm max}} \rangle$ and its
variance for some a range of $x'$ for $\beta_{k} = k^{5/2}$.
Due to scaling behavior $\langle k_{{\rm max}} \rangle/A$
for various $A$'s approach the single curve shown above.

At the critical point, $\langle n_{k} \rangle \sim k^{-\tau}$ where
$\tau > 2$ can be chosen when specifying the original weight.
The critical exponent $\tau$ for percolation is 2.21.

\section{Conclusion and Summary}
\label{sec:Conclusion}

In this paper, hadronic matter under some extreme conditions was
investigated using an exactly solvable model which can describe both
the fragmentation of nuclei and the thermal properties in a unified
way.  Three distinct types of behavior characterize hadronic matter.
At low temperatures, the fermionic aspects of the system are apparent
through such quantities as Fermi degeneracy pressure and Fermi energy.
At high temperature and low density, ideal gas laws are shown to
characterize the system.  Between these two limits, a regime exists
in which the behavior of the fragmenting matter resembles an ideal Bose
gas with its associated Bose-Einstein condensation.  The consequences
of Bose-Einstein condensation on the properties of hadronic matter
were studied.  Simple expressions were obtained for the nuclear
equation of state, heat capacity, compressibility, entropy, energy 
and fragmentation yields which encompass these three regimes.
The importance of multiplicity fluctuations in
statistical models of nuclear fragmentation and its effect on various
thermodynamic properties is stressed.  Some recent data from the EOS
collaboration on critical point behavior of hadronic matter is
analyzed and simple expressions from the exactly solvable
model are obtained which can account for some features of the data.

\acknowledgments
This work supported in part by the National Science Foundation
Grant No. NSFPHY 92-12016 and by the Department of Energy Grant No.
DE-FG02-96ER40987.

\appendix
\section{Multiplicity expectation values}
\label{sec:appendix}

Given a function $f(m)$ where $m$ is the multiplicity,
it is often useful to compute
$x {\partial \over \partial x} \langle f(m(x, y)) \rangle$.
If $f(m)$ is a polynomial, the following identities can be used to
compute $x {\partial f \over \partial x}$
\begin{eqnarray}
x {\partial \over \partial x} \langle [m]_{k} \rangle
  & = & \langle m [m]_{k} \rangle - \langle m \rangle \langle [m]_{k} \rangle
        \;, \nonumber \\
x {\partial \over \partial x} \langle m^{k} \rangle
  & = & \langle m^{k+1} \rangle - \langle m \rangle \langle m^{k} \rangle
        \;, \nonumber \\
x {\partial \over \partial x} \langle m \rangle_{k}
  & = & \langle m \rangle_{k+1} 
    - k \langle m \rangle_{2} \langle m \rangle_{k-1}
        \nonumber \;.
\end{eqnarray}
where $[x]_{n}$ = x(x-1)....(x-n+1) and $\langle m \rangle_{n}$ = nth
cumulant moment of m.
The first is easily derived from the fact that
$\langle [m]_{k} \rangle = {x^{k} \over Z_{A}(x)} {\partial^{k} Z_{A} \over \partial x^{k}}$.  
The second identity can be derived from the first
by expanding
$\langle m^{k} \rangle = \sum_{j=0}^{k} {k \atopwithdelims\{\} j} \langle [m]_{k} \rangle$,
where ${k \atopwithdelims\{\} j}$ is the
Stirling number of the second kind, combined with the
additional identity
$\langle [m]_{k} \rangle = \sum_{j=0}^{k} (-1)^{k-j} {k \atopwithdelims[] j} \langle m^{k} \rangle$,
where ${k \atopwithdelims[] j}$ is the
unsigned Stirling number of the first kind, and the orthogonality
relations between the two Stirling numbers.
The third can be derived from the
second using the binomial expansion of $(m - \langle m \rangle)^{k}$.

These three identities are supplemented by three additional identities
with 
$x {\partial \over \partial x} \rightarrow -y {\partial \over \partial y}$.



\begin{figure}
\caption{Pressure (a) and (in)compressibility (b) of the model at various
temperatures.}
\label{fig:pressure}
\end{figure}

\begin{figure}
\caption{Gibbs free energy per particle of the model as a function of
pressure at various temperatures.}
\label{fig:gibbs}
\end{figure}

\begin{figure}
\caption{Internal energy (a) and specific heat (b) per particle of the model
at various densities.}
\label{fig:spec-heat}
\end{figure}

\begin{figure}
\caption{Entropy per particle of the model at various densities.}
\label{fig:entropy}
\end{figure}

\begin{figure}
\caption{The variance in the nuclear multiplicity of the model as a
function of $x'/x'_{c}$ (a) and $\langle m \rangle/A$ (b).}
\label{fig:mult}
\end{figure}

\begin{figure}
\caption{Size of the largest cluster (a) and its variance (b) of the model as
a function of $x'/x'_{c}$.}
\label{fig:kmax}
\end{figure}

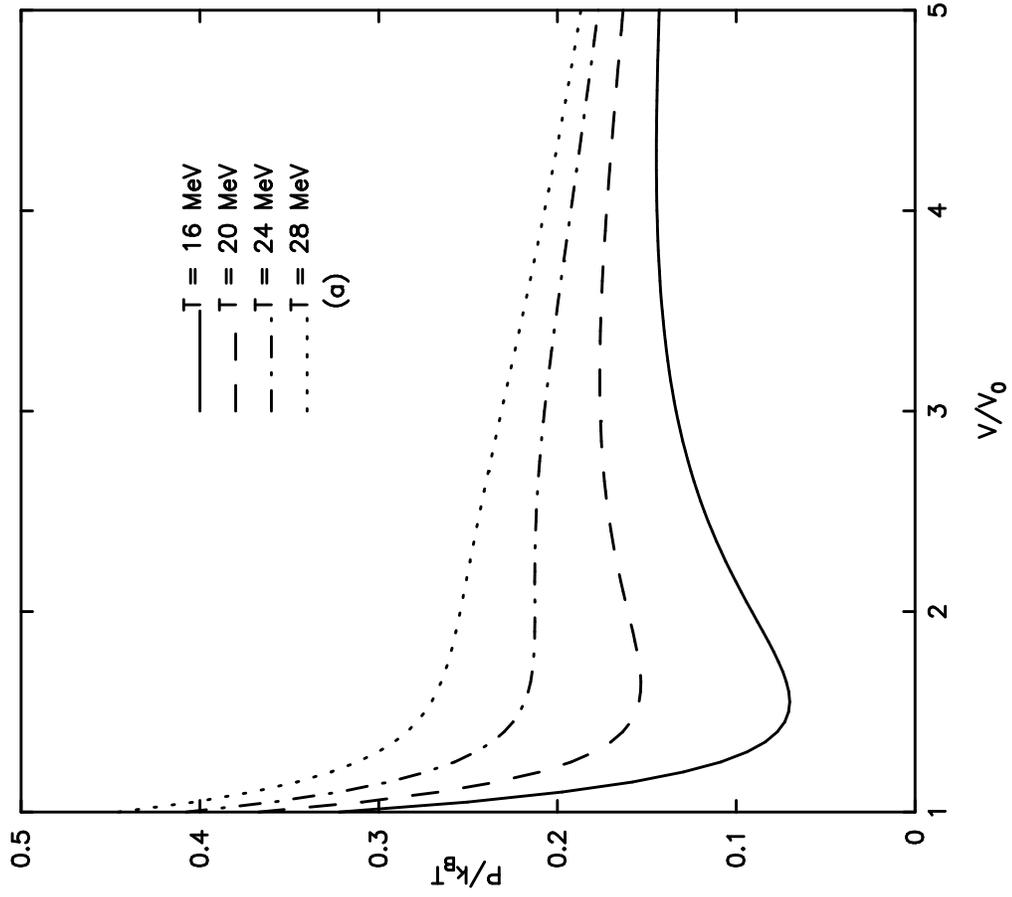
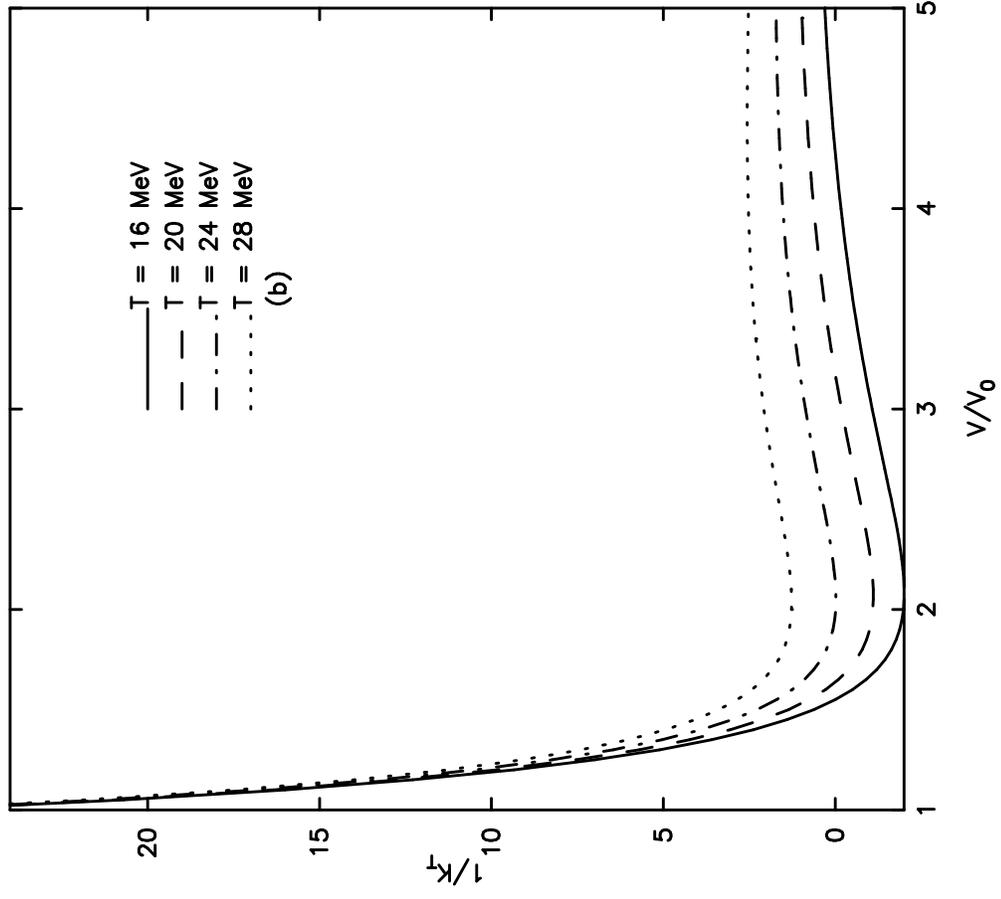
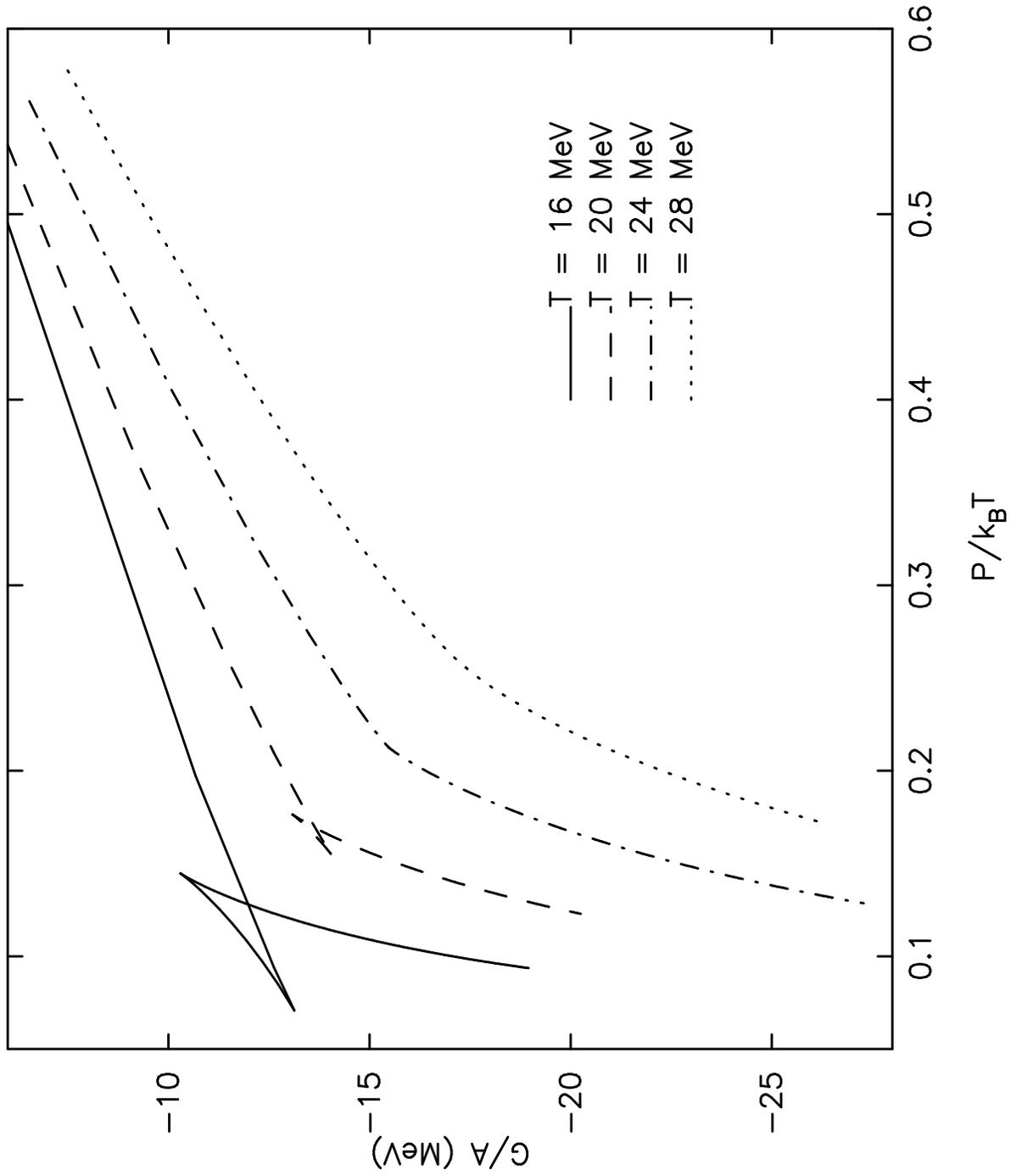
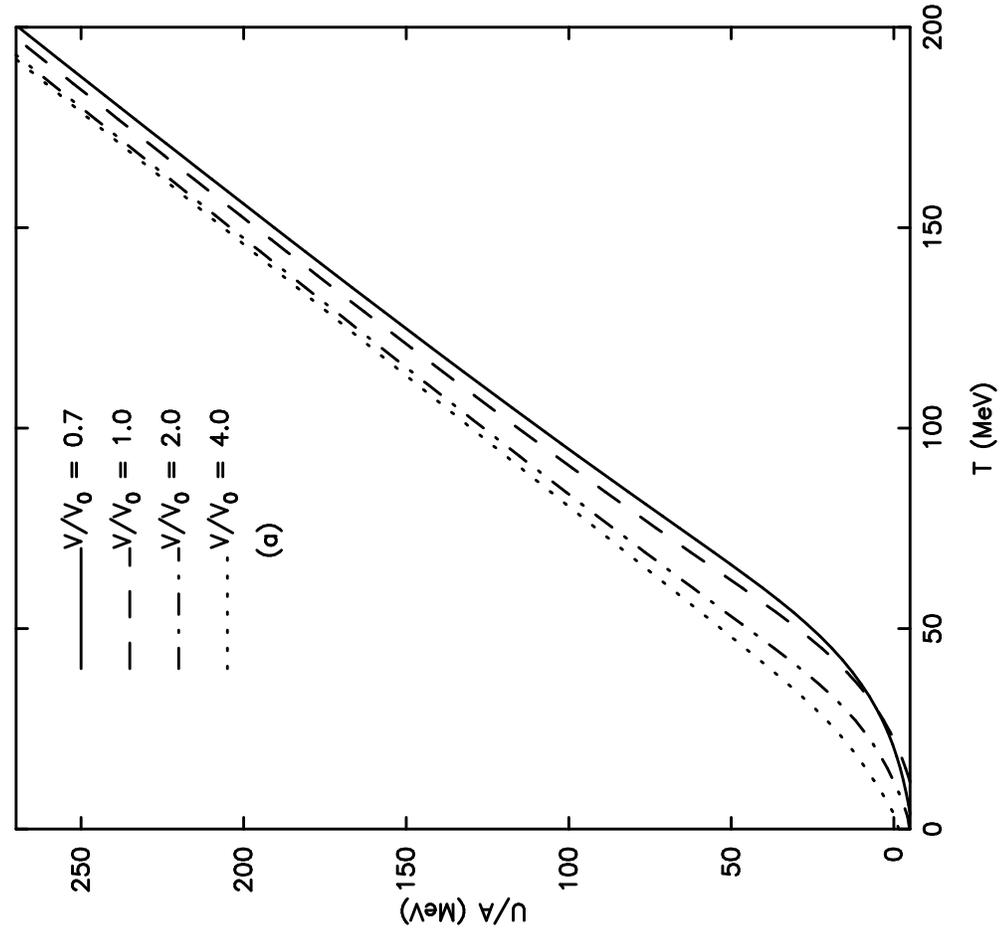
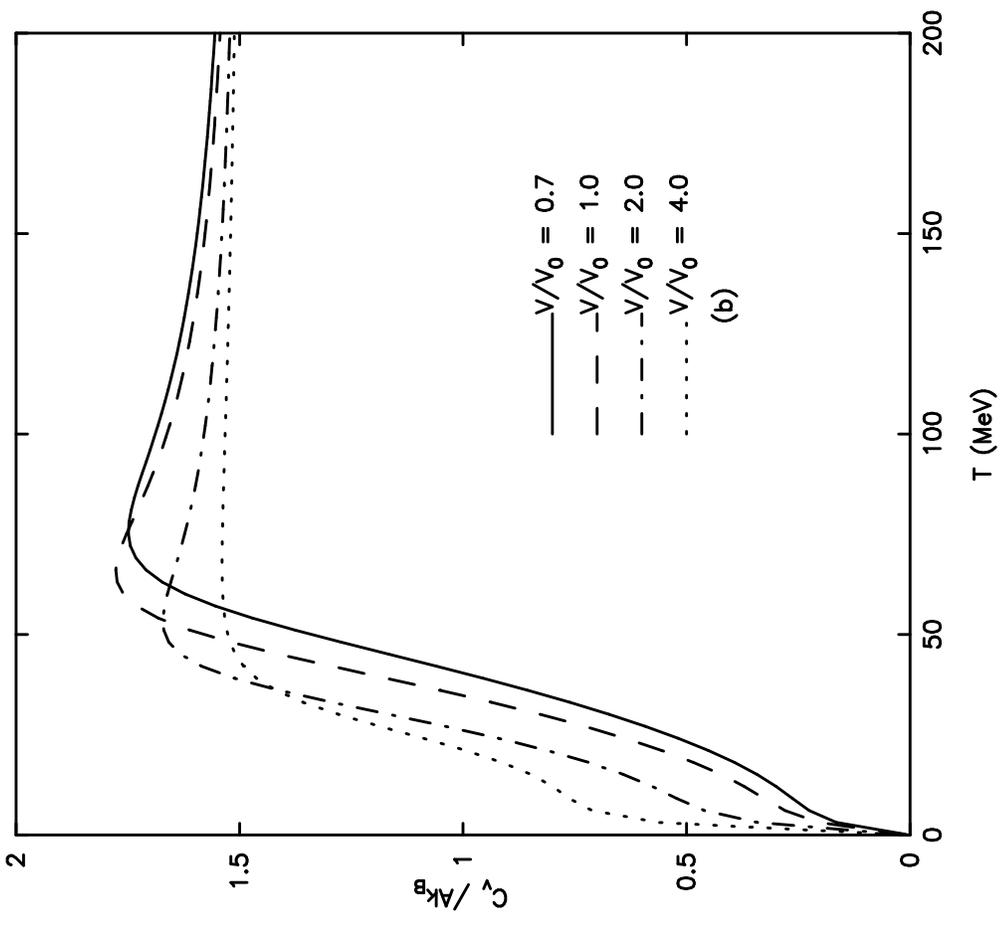
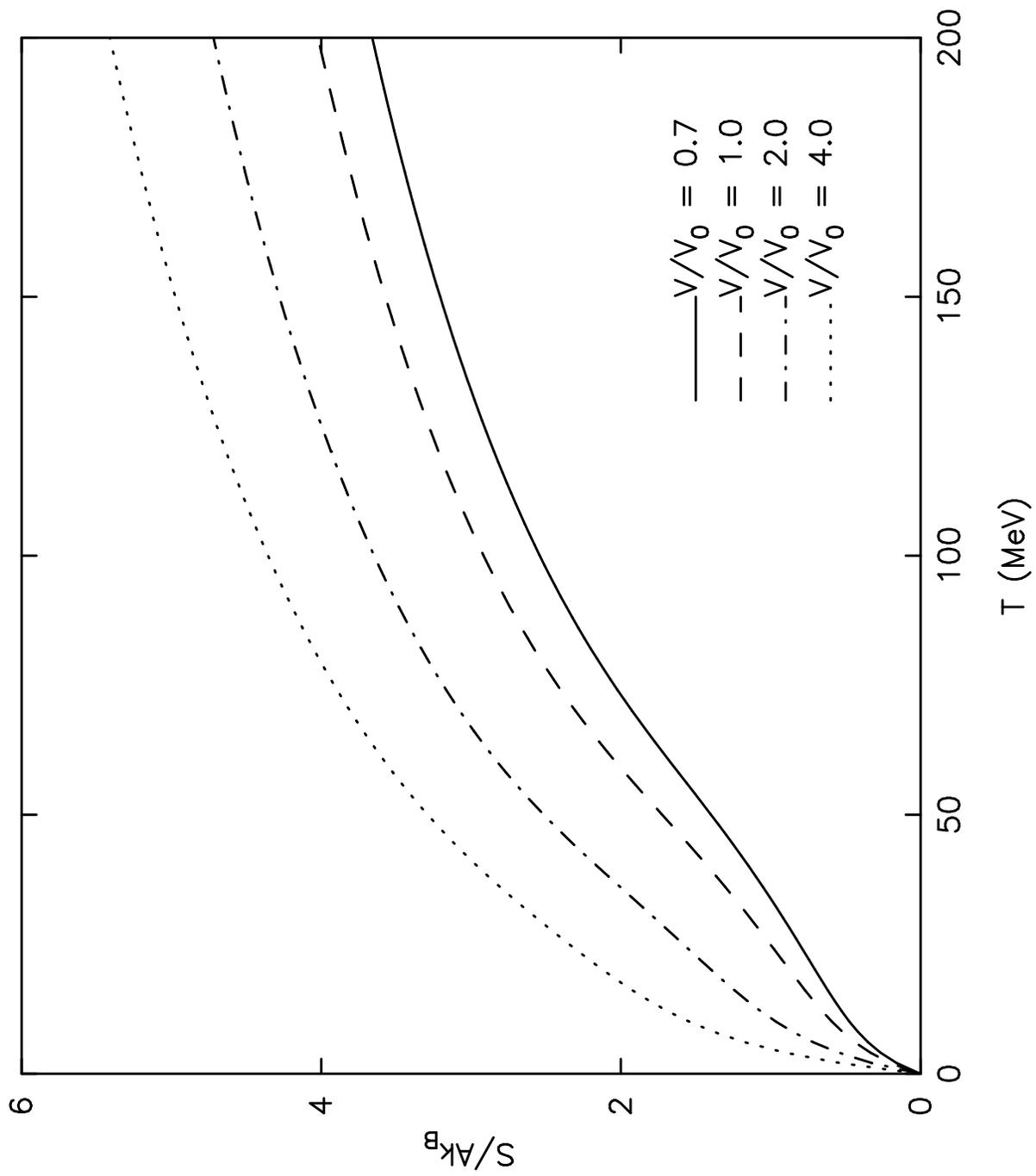
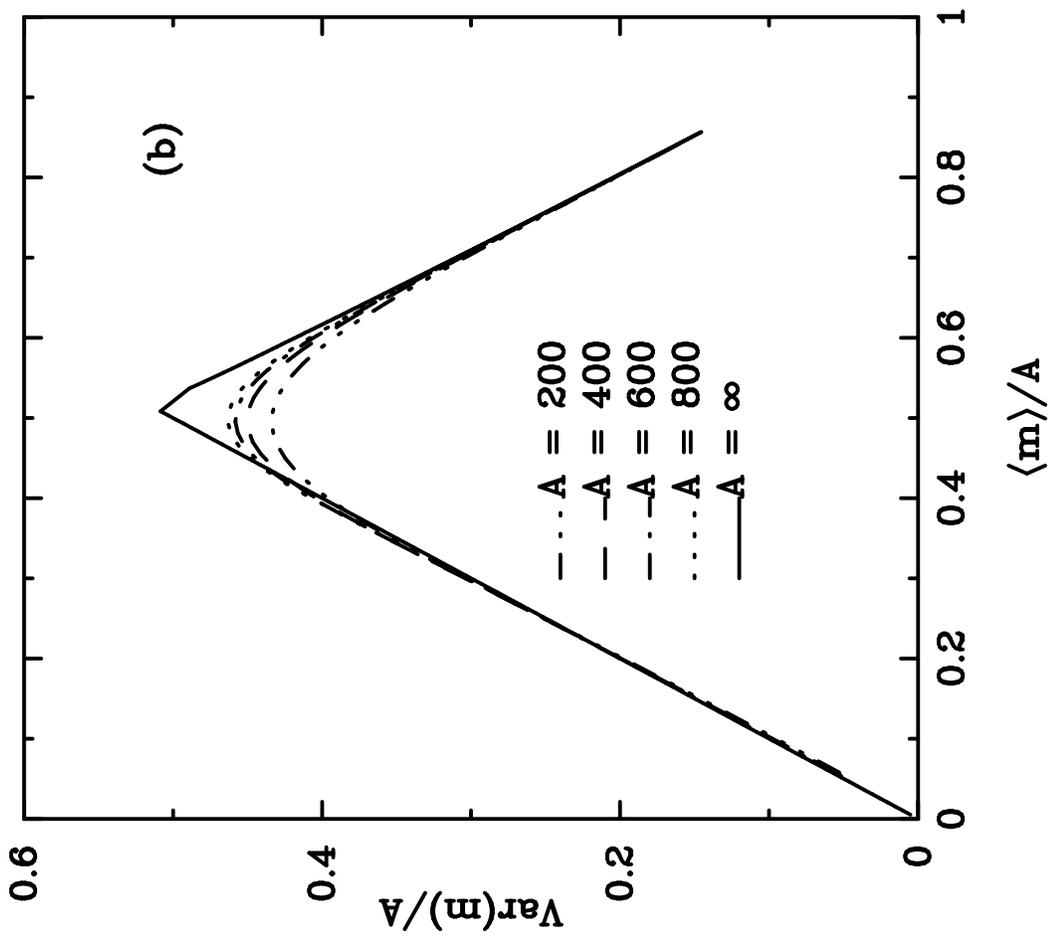
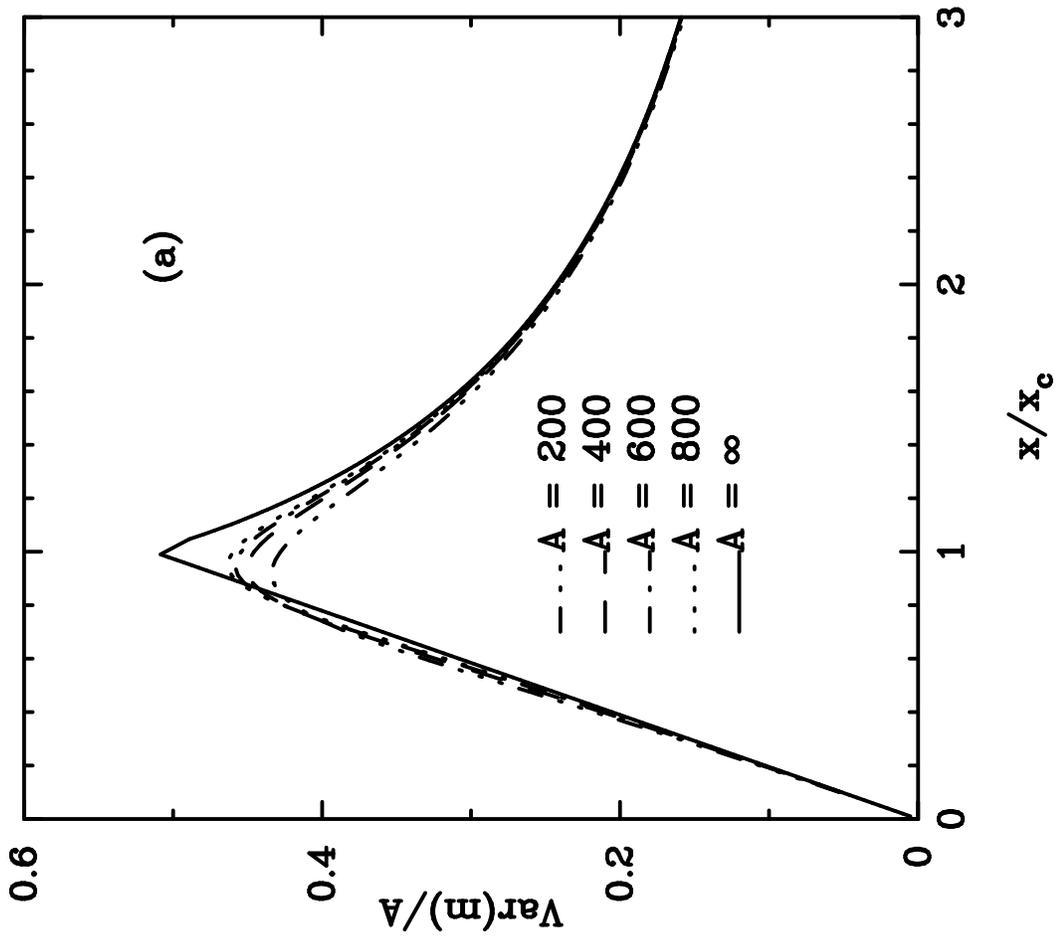
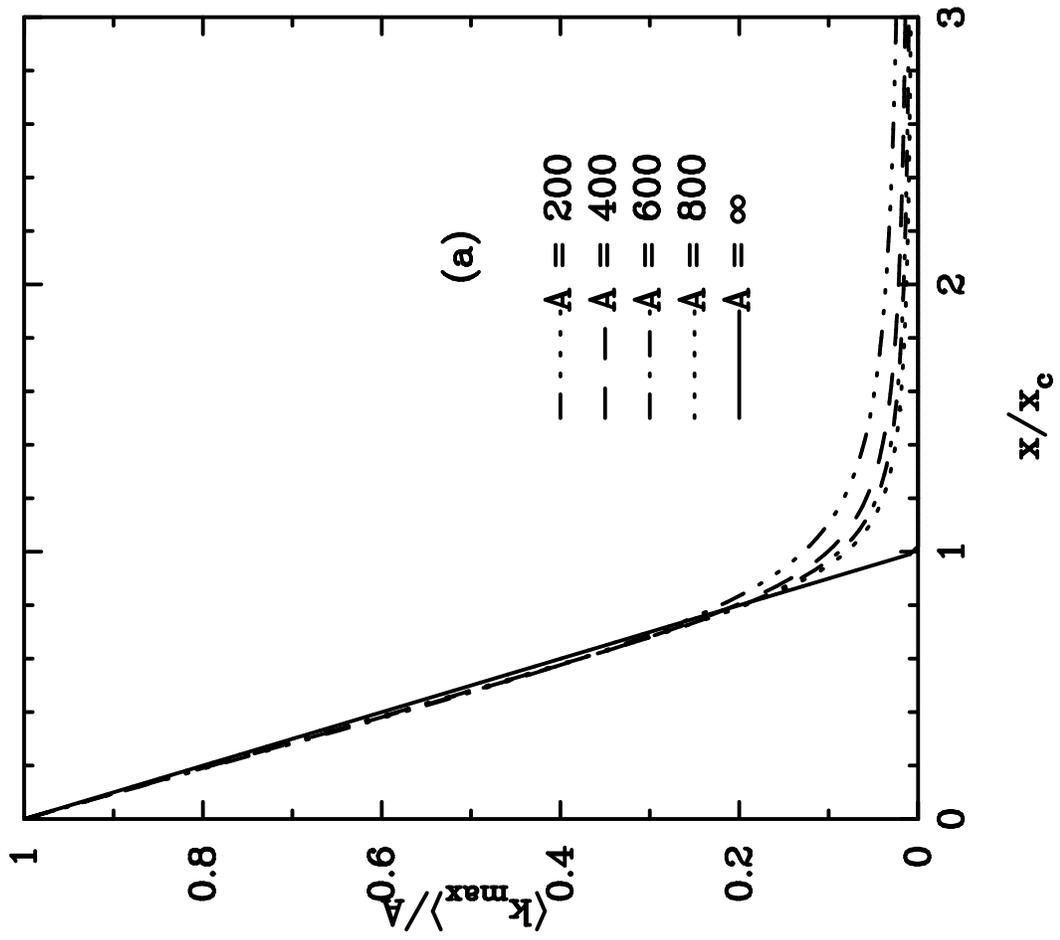
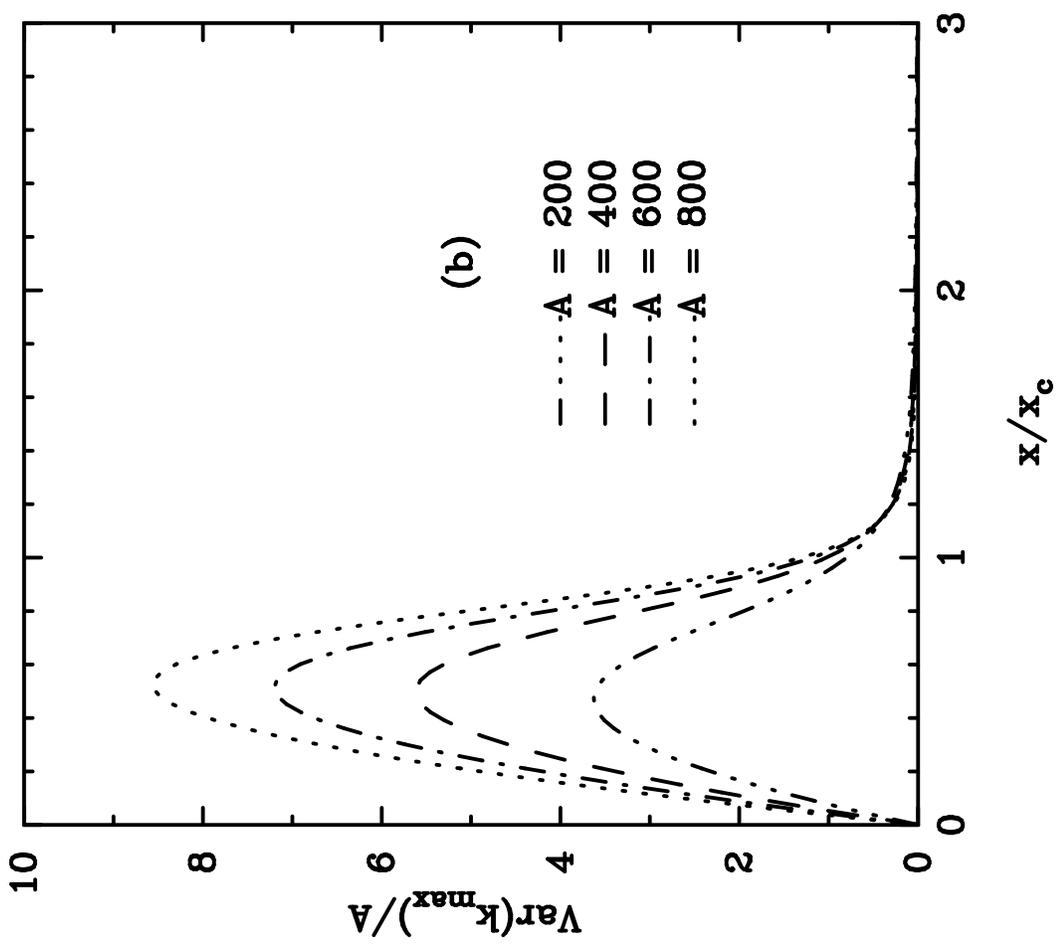
\end{document}